\documentclass[11pt]{article}

\usepackage{epsfig}
\usepackage{latexsym}

\newcommand{\bi}{\begin{itemize}}
\newcommand{\ei}{\end{itemize}}
\newcommand{\be}{\begin{equation}}
\newcommand{\ee}{\end{equation}}

\newcommand{\bea}{\begin{eqnarray}}
\newcommand{\eea}{\end{eqnarray}}
\newcommand{\beastar}{\begin{eqnarray*}}
\newcommand{\eeastar}{\end{eqnarray*}}

\newcommand{\eq}[1]{(\ref{#1})}

\newcommand{\s}{\sigma}
\newcommand{\eps}{\epsilon}

\begin{document}

\title{Resonant Nonequilibrium Temperatures}

\author{F Ritort~$^{\dag}$\\
$\dag$ Department of Physics, Faculty of Physics, University of Barcelona\\ 
Diagonal 647, 08028 Barcelona, Spain\\
{\tt E-Mail:ritort@ffn.ub.es}}

\maketitle

\abstract{In this paper we investigate nonequilibrium temperatures in
a two-state system driven to a nonequilibrium steady state by the
action of an oscillatory field. The nonequilibrium temperature is
determined by coupling a small cavity or probe to the nonequilibrium
system and studying the fluctuating noise in the cavity, as has been
proposed in the context of glassy systems.  We show the presence of
resonant effects in the nonequilibrium temperature and discuss the
existence of a constitutive steady-state equation in such
nonequilibrium conditions. We propose this simple model as an
excellent system to carry out experimental measurements of
nonequilibrium temperatures. This may help to better understand the
physical meaning of such an elusive concept.}

\section{Nonequilibrium Temperatures}
Nonequilibrium physics is a vast field of research plagued by many
interesting open questions. A question that has attracted the
attention of theorists for a long time is the possibility to define a
nonequilibrium temperature$^{1}$ (NET).  There is an important motivation behind the quest for such
concept. One of the most important conceptual ingredients in classical
thermodynamics is the notion of temperature. Temperature is a physical
measure of the energy content of a system. When two bodies at
different temperatures are put in contact with each other, heat flows
from the body at higher temperature to the body at a lower
temperature. This experimental fact allows the classification of thermodynamic
states according to the value of their temperature (defined after
establishing a suitable temperature scale). Bodies at identical
temperatures do not exchange a net energy current when put in thermal
contact.

The main goal when extending the concept of temperature to
nonequilibrium systems is to establish under which conditions the
previous experimental facts are still valid. But what apparently seems
to be an easy promenade along the foothills of a new land turns out to
be a fatal excursion full of traps and difficulties. It is not our
purpose to enumerate here the casuistry of examples of nonequilibrium
systems (many of them pretty simple and harmless) where such an extension
has proved unsuccessful. Most of the difficulty lies in the
transitivity property of temperature as well as in the fact that
nonequilibrium bodies can be put in contact in many different ways,
which do not necessarily lead to the same result.  Yet, there is an
important exception to all this complicated phenomenology, put forward
by Onsager a long time ago. We refer to irreversible or nonequilibrium
thermodynamics$^2$ in the regime where deviations from
equilibrium are not large enough for currents (of heat, mass, energy,
or charge) to be proportional to the externally applied forces. Sometimes
this is also referred to as the linear response regime where fluctuations
are Gaussian distributed. Linear systems are pretty interesting as
they teach us that the NET must be a local
concept$^3$. For instance, in a wire of metal with the
extremes in contact with two thermal sources, heat flows form the
higher temperature source to the lower temperature one. The heat
current is then proportional to the temperature gradient, and a
temperature field (e.g. varying linearly along the wire) describes the
nonequilibrium state of the system.

However, there is a more recent category of systems that deserve
special attention because they seem to be good examples to investigate
the concept of NET. These are aging or glassy systems that are
prepared in an initial far-from equilibrium state and slowly evolve
toward equilibrium for astronomically large time scales. Under such
conditions a statistical interpretation of fluctuations can be
quantitatively described by an extended fluctuation-dissipation theorem (FDT)
introduced in the context of spin glasses$^{4,5}$. As
these systems slowly evolve toward equilibrium they appear to
thermalize over some regions of phase space and develop a restricted
flat measure (we may refer to it as a {\em protomeasure}) similar to
the microcanonical measure for equilibrium systems. The flat state
reached by the system under such conditions defines a statistical
effective temperature similar to that provided by statistical
mechanics$^{6,7}$.  One important aspect of this NET is its dependence on
the probed frequency as well as the age of the system. It has been
suggested by Cugliandolo, Kurchan, and Peliti that such a pattern of NETs
could be experimentally measured by using a small interacting probe
that plays the role of a thermometer whose frequency can be
tuned$^8$. In the glassy jargon, it is common to denote this NET as
effective temperature.

\section{Motivation}
Few experimental measurements of the NET exist. To date, the
interpretation of most of these results still remains unclear. The
first indirect experimental measure of NETs in aging systems dates
back to magnetic noise measurements in spin glasses$^9$. In these
measurements systematic deviations for the FDT were not observed, yet
these did not probe the relevant frequency range. A few years ago,
Israeloff and Grigera did voltage noise measurements in an electric
resonant circuit formed by a capacitor containing glycerol and an
inductance$^{10}$. Such experiments showed a faint dependence of the
effective temperature on the age of the system. Subsequent Nyquist
noise measurements by Ciliberto and co-workers on Laponite and
polycarbonate have shown that the effective temperature is strongly
dependent on frequency$^{11}$. At low enough frequencies ($\omega
t_w<<1$, where $t_w$ denotes the age of the system, also called
waiting time), the effective temperature $T_{\rm eff}(\omega,t_w)$ is
up to 3 or 4 orders of magnitude larger than the bath
temperature. Magnetization response and correlation measurements by
the Ocio group in Paris vindicated strong FDT violations in
spin glasses, and indeed these provided the first experimental
determination of a fluctuation-dissipation plot.$^{12}$ More recently, the measure of NETs
has shifted toward simpler systems and probes such as the stochastic
motion of a torsional pendulum immersed in a granular media$^{13}$ or
a bead joggling in a turbulent wind flow.$^{14}$  More intuitive
results have been obtained in such cases. Optical tweezer measurements
also recently have been used to study the fluctuating motion of a
micrometer-sized bead immersed in an aging Laponite
suspension.$^{15}$ No measurement of the effective temperature has up
to date provided a crystal-clear result of what theory predicts. There
are fundamental questions about the effective temperature about which
we do not yet know the answer. For instance, how is the measured
temperature expected to depend on the characteristics of the probe
(mass, frictional viscosity or size)?  Will the effective temperature
depend on the type of observable measured$^{16}$?. Is the effective
temperature a local or a global quantity?  Ultimately, are the results
reproducible when changing the experimental technique?

The situation is confusing and has some resemblance to other hot
debates currently fashionable in the area of glasses. For example,
regarding the question of heterogeneities,$^{17}$ are
heterogeneities well-defined objects? Is there just one heterogeneity
length scale or are there many? Does the measurement depend on the
experimental technique?

The large amount of unanswered questions and the complexity of the
problem demands going to simpler systems where specific predictions
could be experimentally tested in a systematic way.  In this paper we
propose the study of NETs in two-states driven systems as a useful
playground to elucidate many of the subtleties behind the concept of the
NET. Several reasons compel us to consider such an illustrative example. On
one hand, two-state systems represent the simplest examples of
nonlinear systems with activated behaviors. On the other hand, they are
physically realizable in many different examples in the laboratory and
are easy to control. Finally, steady state systems provide an
interesting nonequilibrium regime characterized by non-Gibbsian
distributions where several theoretical predictions can be tested.  Our
main goal here is to show that, in such conditions, NETs display
resonant effects that are amenable to experimentation.  The present
study is by no means exhaustive; we just want to emphasize the
importance of investigating simpler nonequilibrium systems and
stimulate further experiments in this exciting area of research. The
plan of the paper is as follows. In section~\ref{model}, we describe the
model for the nonequilibrium measurements. It consists of a
nonequilibrium system plus an external probe or cavity used for the
measurements. In Sec.~\ref{results} we show the main results: existence
of resonant effects in the NET and the possibility of defining a
constitutive equation for the steady state. Finally, we discuss candidate
systems where these effects could be experimentally tested.

\section{Model}
\label{model}
An interesting problem in statistical physics is to know whether it is
possible to extend some of the concepts of thermodynamics to
nonequilibrium systems in steady states (SS)
and possibly characterize them by nonequilibrium constitutive
equations.$^{18}$ Nyquist noise measurements offer many
potentialities as these are an ideal tool to probe thermal
fluctuations around the steady state. Heat fluctuations are
experimentally accessible by using a resonant cavity coupled to the
system. Although such measurements have been attempted in different
contexts, experimental verifications are scarce, probably because of
the great complexity of the systems addressed,$^{10,11}$ and
experimental measurements turn out to be difficult. In this regard,
numerical simulations become an alternative tool of
research. Recently, Hatano and Jou$^{19}$ have investigated
NETs in mechanically driven linear oscillators by coupling them to a
resonant cavity. Here, we propose measurements of NETs in two-state
systems (e.g., magnetic-nanoparticle or electric-dipole systems)
coupled to a resonant probe and show the existence of resonant effects
that are modulated by the physical properties of the cavity or
probe. At the end, we suggest possible experiments that could check
such predictions.

\subsection{The system}

A broad category of systems can be modeled by two-level systems. These
offer realistic descriptions of electronic and optical devices that can
function in two different configurations, atoms in their ground and
excited states, magnetic particles whose magnetic moment can point in
two directions, or biomolecules in their native and unfolded states,
among others. In what follows, and for sake of clarity, we will adopt the nomenclature of magnetic
systems.  A two-state unit has the magnetic moment $\mu$ and can point in
two directions according to the sign of the spin $\s=\pm 1$. In the
presence of an external field $H$, the energy of the spin is given by
$E(\s)=-\mu H \s$. The transition rate for the spin will
be denoted as $p^{\rm up}(H)$ and $p^{\rm down}(H)$ to indicate the
transitions $\s=-1\to\s'=1$ and $\s=1\to\s'=-1$, respectively.  These
rates satisfy detailed balance; therefore, $p^{\rm up}(H)/p^{\rm
down}(H)=\exp(-2\beta \mu H)$ where $\beta=1/k_BT$, with $T$ being the bath
temperature and $k_B$ being the Boltzmann constant. The overall transition
rate is given by $p^{\rm tot}(H)=p^{\rm up}(H)+p^{\rm
down}(H)$. We have chosen Glauber transition rates given by
\be
p^{\rm up}(H)=p^{\rm tot}(H)q(H)~~~;~~~p^{\rm down}(H)=p^{\rm
tot}(H)(1-q(H))
\label{num1}
\ee
with $q(H)=(1+\tanh(\beta\mu H))/2$ and $p^{\rm
tot}(H)=1/\tau_{\rm relax}(H)=\alpha(H)$ corresponding to the inverse of the
relaxation time. 

For the nonequilibrium state, we just consider that the spin is driven
by an oscillating field of frequency $\omega$ of the form
$H(t)=H_0\cos(\omega t)$. The transition rates for the spin are now time
dependent and given by,
\bea
p^{\rm up}(t)=\alpha(H(t)) \frac{\exp(\beta\mu H(t))}{2\cosh(\beta\mu H(t))}\label{num2a}\\
p^{\rm down}(t)=\alpha(H(t)) \frac{\exp(-\beta\mu H(t))}{2\cosh(\beta\mu
H(t))}~~~~.
\label{num2}
\eea
For the relaxation time $\tau_{\rm relax}(H)$ or $\alpha(H)$ we have
chosen the barrier dependence mostly found in  thermally activated magnetic
systems (such as magnetic nanoparticles$^{20}$). These are described by the
Brown-Neel formula,
\be
\tau_{\rm relax}(H)=\tau_0\exp\Bigl( \frac{B(H)}{k_BT} \Bigr) 
\label{eq1}
\ee
where $\tau_0$ is a microscopic time describing relaxation within a
state and $B(H)$ is a field-dependent activation barrier. Related
expressions exist for the case of a molecular bond that breaks under
the action of a mechanical force.$^{21}$ In that case, the spin
corresponds to the extension of the molecule, and the magnetic field
is the applied force at the ends of the molecule. In the present
study, we have considered the simplest case where $B(H)=B_0$ is
field-independent.  This is a reasonable approximation. In
superparamagnetic nanoparticles, the energy barrier is nearly
field-independent (in contrast to ferromagnetic nanoparticles with
uniaxial anisotropy where $B(H)$ is not constant but depends on the
intensity of the external field projected on the easy magnetization
axis). Superparamagnetic nanoparticles can be experimentally realized
in paramagnetic molecular clusters without a coercitivity field (yet,
for these systems the magnetic signal of individual nanoparticles is
expected to be very small).  Other examples are specific ferro- and
ferrimagnetic nanoparticles where the anisotropy contribution to the
zero-field barrier is negligible (for a discussion see ref ${22}$).
The height of the barrier usually depends on the properties of the
nanoparticle. Recent experiments have demonstrated how the height of
the barrier $B_0$ can be reduced (therefore $\tau_{\rm relax}(H)$ can
be modulated) making it possible to observe magnetization-reversible
transitions (also called telegraph noise measurements) in single
cobalt nanoparticles at low temperatures.$^{23,24}$

\subsection{The cavity}
We model the resonant cavity as a harmonic oscillator with natural
frequency $\omega_0$, mass $m$, and friction coefficient $\gamma$ (for
an electrical cavity these correspond to a capacitance, an inductance,
and a resistance, respectively) coupled to a bath at temperature
$T'$. These parameters define an inertial time scale
$\tau_1=2\pi/\omega_0$ and a dissipation time scale $\tau_{\rm
dis}=\frac{\gamma}{k}$ where $k=m\omega_0^2$ is the stiffness constant
of the cavity.  To model the interaction between the spin and the
cavity we add a term in the energy of the type $-\eps\s x$ in $E(\s)$,
$E(\s)=-\mu H \s-\eps\s x$ as well as in the equation of motion of the
oscillator, where $x$ describes the observable associated to the
cavity (e.g. the voltage or charge of the capacitor)~\footnote{This
coupling can be interpreted in many ways. For instance, $\s$ could
stand for the magnetic ($\vec{m}$) or electric ($\vec{p}$) dipole moment of the system and $x$ for the
magnetic field (generated by an inductance) or the electric field
(generated by a capacitor) of an electrical cavity.  Therefore, the coupling would be linear of the form
$\vec{m}\vec{B}$ or $\vec{p}\vec{E}$, respectively. Moreover, to a first
approximation, the magnetic field is proportional to the
current intensity $I$ and the electric field is
proportional to the voltage across the plates of the capacitor
$V$. Therefore, $x$ stands for either the current intensity $I$ or
voltage $V$ (or charge $Q$) of the capacitor.}. The cavity is governed by
the stochastic equation
\be
m\ddot{x}+\gamma\dot{x}+m\omega_0^2 x-\eps\sigma=\xi
\label{eq2}
\ee
where $\xi$ is a white noise with correlations
$\langle\xi(t)\xi(s)\rangle=2k_B T'\gamma\delta(t-s)$. Let the
magnetic system be driven to a SS under the action of an oscillating
external field $H(t)=H_0\cos(\omega t)$, and let us put it in contact
with the cavity as described previously. The average power supplied by
the system upon the cavity is given by ${\cal P}=\eps\overline{\sigma
\dot{x}}$ where the bar over the characters stands for the average
over many cycles. If $T'\gg T$, then an average heat current will flow
from the cavity to the system, and if $T'\ll T$, then the contrary
will hold. For small enough $\eps$, the average power transferred
between the probe and the system will be proportional to
$\eps^2$. This result can be simply understood from symmetry
considerations. By a change of $\eps\to-\eps$ in eq \eq{eq2} the
dynamics of the probe are invariant under the spin inversion
$\s\to-\s$. This last transformation also does not change the dynamics
of the spin in the ac field. Hence, the transferred power ${\cal P}$
must remain invariant under the transformation $\eps\to-\eps$. Upon
variation of $T'$, there must be a temperature
$T_m=T'_{\omega,\omega_0,m,\gamma}(T)$ (which depends on the
parameters of the cavity) at which the average power supplied to the
cavity vanishes in a quadratic order in $\epsilon$,
i.e. $\overline{\sigma \dot{x}}=0$ in a linear order in
$\eps$. Sometimes the net heat power exchanged between the system and
the thermometer is considered, ${\cal P}=\eps(\overline{\sigma
\dot{x}-x\dot{\sigma}})$ (where we take
$\dot{\sigma}(t)=\sigma(t+1)-\sigma(t)$). Note that, from the relation
$\dot{x\sigma}=\dot{x}\sigma+x\dot{\sigma}$, there is no difference
between this and our definition as they coincide (except by a global
sign) for SS systems. Only for non-SS systems (e.g., aging systems),
would we expect to see differences between both quantities$^{25}$.

\section{Results}
\label{results}
We have numerically integrated the stochastic equations eqs \eq{num2} and \eq{eq2}
under different conditions. The dynamics of the spin (eq \eq{num2}) have been
simulated by taking into account its interaction with the
cavity. Note that, when solving the coupled dynamics, the system and the cavity feel the ${\cal O}(\epsilon)$
interaction with their corresponding partner. Both the interaction of
the system with the cavity and viceversa contribute to the total
transferred power ${\cal P}$ in the second order of $\eps$.

The number of parameters that we can vary in the simulations is very
large. At first, there seem to be four different time scales $\tau_{\rm
relax},1/\omega,\tau_{\rm dis}$ and $1/\omega_0$. In what follows, we show
results for a fixed value of the dissipative time scale of the cavity
$\tau_{\rm dis}=\frac{\gamma}{k}$. However, out of the three remaining
time scales, we are only left with just two independent variables. The
argument is as follows. For a given value of the frequency of the ac
driving field $\omega$ and a value of the relaxation time of the spin
$\tau_{\rm relax}$ (eq \eq{eq2}), the results only depend on the adimensional
product $\omega\tau_{\rm relax}$ as one time of the two time scales
($1/\omega$ or $\tau_{\rm relax}$) can be used to rescale the time in
the equation of motion for the spin in the SS.  For a fixed value of
$\omega\tau_{\rm relax}$, the value of $T_m$ depends only on the
ratio $\omega_0/\omega$ between the frequency of the oscillator cavity
$\omega_0$ and the ac driving field $\omega$. Again, another time scale
($\omega_0$ or $\omega$) can be rescaled in the equation of motion for
the oscillator.  Therefore, after having fixed the value of $\tau_{\rm
dis}$ we are just left with the two adimensional quantities
$\omega\tau_{\rm relax}$ and $\omega_0/\omega$ as free parameters.  For each
pair of the values $\omega\tau_{\rm relax}$ and $\omega_0/\omega$ we have
determined the value of $T_m$ for which the average power supplied to the
cavity vanishes.  The results obtained are shown in Figure~\ref{fig1} as
the ratio $\omega_0/\omega$ changes by several orders of magnitude (each
curve corresponding to a value of $\omega\tau_{\rm relax}$).  The
following parameters were chosen for the two-states system: $ \tau_{\rm
dis}=1s,\tau_0=10^{-7}s,B_0=2300K$ and $T=200K$ corresponding to a
relaxation time of the spin $\tau_{\rm relax}=0.01s$ as given in
eq \eq{eq1}.

We note several salient features in this figure:

\begin{enumerate}

\item{\bf Resonance effect.} There is a resonant peak in $T_m/T$ for
$\omega=\omega_0$ that gets more sharp as the equilibrium regime
$\omega\tau_{\rm relax}\to 0$ is approached. On the contrary, if
$\omega\tau_{\rm relax}$ increases, then the resonance effect fades
away. This effect can be easily understood. When $\omega\tau_{\rm
relax}\to 0$ the system is driven slowly enough for the driven process
to be quasi static or reversible. In this regime, the amount of
dissipated heat (defined as the average dissipated work$^{26}$ along a cycle of
the field) decreases proportionally to the ratio $\omega\tau_{\rm
relax}$, and mechanical resonance is observed.
\begin{table}[b]
\vspace{-0.cm}
\begin{center}
\begin{tabular}{|c||c|c|c|c|c|c|c|c|c|}\hline
$\omega\tau_{\rm relax}$&\multicolumn{1}{c}{0.001}\vline&\multicolumn{1}{c}{0.01}\vline&\multicolumn{1}{c}{0.02}\vline&\multicolumn{1}{c}{0.1}\vline&\multicolumn{1}{c}{1}\vline\\\hline\hline
$\alpha$&1.74&2.48  &2.81 &0.15 & 0.14\\\hline
$\beta$ & 0 & 3.33& 3.24& 0.25& -- \\\hline
$a$ &1.79 &1.68  &2.10 &0.98 &2.87 \\\hline
$b$ &1.04 & 2.59 &2.88 &5.45 & -0.02\\\hline
$c$ &0.30 &0.76  &1.21 &2.25 & -- \\\hline
\end{tabular}
 \caption{Parameters for the Fit to the Spectral Line Shape
 \eq{eq3}. For the largest value of $\omega\tau_{\rm relax}=1$, data
 have been fitted to the function
 $T_m(\tilde{\omega})/T=a+b\tilde{\omega}^{-2\alpha}$. For
 $\omega\tau_{\rm relax}=0.1$, the results of the fit are only
 approximate (see the discussion in footnote~\ref{foot2}).}
\label{table}
 \end{center}

\end{table}

\item{\bf Spectral line shape.} The value of $T_m/T$ becomes
independent of the frequency of the cavity for large enough values of
$\omega_0$.  $T_m$ values can be well fitted to a spectral line shape of the form
\be
T_m(\tilde{\omega})/T=(a+b\tilde{\omega}^{2\alpha})/((\tilde{\omega}^{\alpha}-1)^2+c\tilde{\omega}^{\beta})
\label{eq3}
\ee
with $\tilde{\omega}=\omega_0/\omega$ and $\alpha$ and $\beta$
exponent values satisfying $\beta<2\alpha$. The fit works well for not
too large values of $\omega\tau_{\rm relax}$ when the resonance peak
is observed.  For large values of $\omega\tau_{\rm relax}$
(Figure~\ref{fig1}), the curve $T_m(\tilde{\omega})/T$ becomes flat
(no resonance peak is observed), and the fit is only an approximated
guess.  In the limit $\omega\to 0$, the right side of the spectral line
shape converges to a Lorentzian plus a constant (equal to 1). The
horizontal dot-dashed line $T_m/T=1$ corresponds to the equilibrium
measurement that is only achieved for $\tilde{\omega}\gg 1$ and
$\omega\to 0$.

\item{\bf Nonequilibrium temperature (NET) $T_m^*$.} 
A unique value of $T_m/T$ describes the SS at all frequencies (asterisks
in the figure) corresponding to the strongly dissipative regime where $\omega\tau_{\rm
relax}$ is very large. Moreover, the quantity
\be
T_m^*=\lim_{\omega_0\to\infty}T_m(\omega_0)
\label{eq4}
\ee
has been shown (for instance in the case of weak
turbulence$^{27}$ or in slow relaxing
systems$^8$) to satisfy the Nyquist formula
\be
T_m^*=\lim_{\omega_0\to\infty} \frac{\pi}{2}
\frac{\omega_0\tilde{C}_{\rm cavity}(\omega_0)}{\hat{\chi}_{\rm
cavity}(\omega_0)}
\label{eq5}
\ee
where $\tilde{C}_{\rm cavity}$ and $\hat{\chi}_{\rm cavity}$ stand for
the power spectrum and the out-of-phase susceptibility of the cavity,
respectively. The value of $T_m^*$ is therefore experimentally
measurable by measuring the power spectrum of the
cavity~\footnote{\label{foot2}According to the fit formula (eq
\eq{eq3}), the value of $T_m^*/T$ should be equal to $b$ in the limit
$\tilde{\omega}\to\infty$. However, by comparing the results of the
fit shown in Table~\ref{table} with the results of Fig.~\ref{fig2}, we
find qualitative agreement only for $\omega\tau_{\rm relax}\ll 0.1$
when the resonance peak is observed (for $\omega\tau_{\rm relax}=0.1$,
the value of $b$ is 5.45, too far from the expected value of
approximately 2.8).  This suggests that eq \eq{eq3} is just an
approximate fit to the data shown in Fig.~\ref{fig1}.}.

\begin{figure}
\begin{center}
\epsfig{file=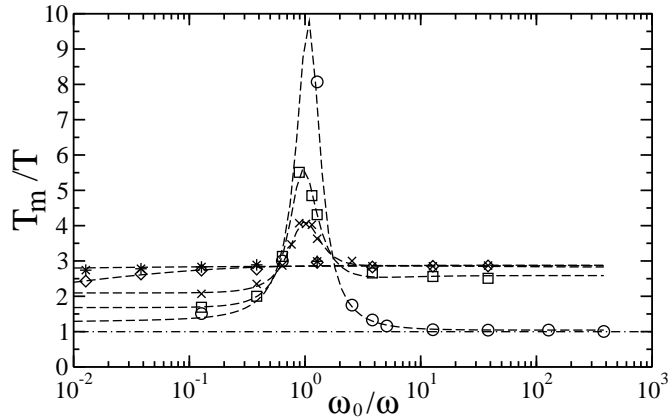,angle=-90, width=10cm}
\end{center}
\vspace{0cm}
\caption{Measured values of the NET, $T_m/T$, for paramagnetic nanoparticles
driven in an oscillating field. The parameters of the model are $\mu=300
\mu_B,T=200K,H_0=2{\rm T},\tau_0=10^{-7}s,$ and $B_0=2300K$. The NET
is determined by coupling the system to a cavity and finding the
temperature of the cavity at which the average heat current between
system and cavity vanishes. The cavity is characterized by different
values of the resonant frequency $\omega_0$ in the range $0.01-100$
rad/s and a damping time scale $\tau_{\rm
dis}=\frac{\gamma}{k}=1s$. Different curves correspond to oscillating
magnetic fields of intensity $H_0=2{\rm T}$ and different frequencies
$\omega\tau_{\rm relax}=0.001 {\rm (circles)},0.01 {\rm (squares)} ,0.02
{\rm (crosses)},0.1 {\rm (diamonds)},1 {\rm (asterisks)}$. The lines are
the spectral line shape fits (see text).}
\label{fig1}
\end{figure}

\end{enumerate}

\subsection{Constitutive Steady-State Equation.} 
\begin{figure}
\begin{center}
\epsfig{file=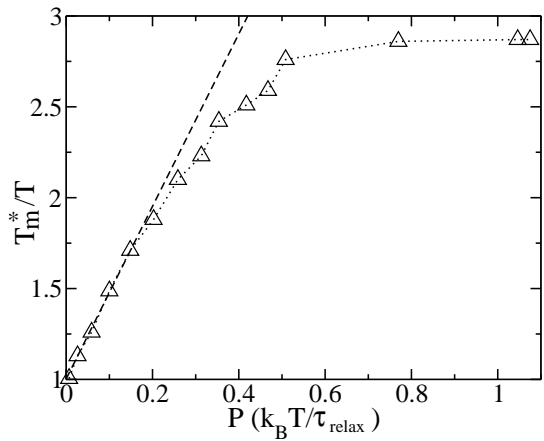,angle=-90, width=9cm}
\end{center}
\vspace{0cm}
\caption{Constitutive equation of the magnetic system in its SS. $P$
stands for the entropy production per particle of the magnetic
system. The entropy production is expressed as the dissipated heat or
work (the area enclosed in the $M-H$ hysteresis cycle in $k_BT$ units)
divided by the relaxation time of the particles $\tau_{\rm relax}$ as
given in eq \eq{eq1}. The dashed straight line is the linear fit
$P=\alpha(T_m^*/T-1)$ with $\alpha\simeq 0.2$. The dotted line passing
through the points is a guide to the eye. The same parameters are used as in
Figure~\ref{fig1}.}
\label{fig2}
\end{figure}

The NET $T_m^*$ was determined in numerical simulations of a forced
linear oscillator and shown to be dependent on the type of coupling
between system and cavity$^{19}$. For the two-states or spin
system we have verified this result by changing the type of coupling
between oscillator and system. When the coupling energy term is changed
from $-\eps\sigma x$ to $-\eps\sigma x^2$ the value of $T_m$ changes
noticeably (data not shown).  Yet the value of $T_m^*$ can be a useful
quantity to characterize the SS. As we said previously, the spin
dissipates heat as it tries to follow the oscillations of the field. The
amount of heat lost increases as the driving frequency $\omega$
increases. The amount of heat is given by the area enclosed in the
hysteresis cycle $M-H$ of the system. A quantity that characterizes the
nonequilibrium state is the entropy production $P$ (i.e., the average
dissipated heat per unit time).  In extended irreversible thermodynamic
theories,$^{28}$ the entropy production is expected to be a
function of the previous NET, $T_m^*$. This relation defines a
constitutive steady-state equation. However, at difference with
classical thermodynamics, this relation is expected to have some degree
of ambiguity. Indeed, we have seen that the value of $T_m$ depends on
the types of coupling, therefore a unique relationship (cavity
independent) between the entropy production $P$ and $T_m^*$ cannot be
established in this simple example for all type of couplings. In
Figure \ref{fig2}, we show the SS constitutive equation found in our model;
the value of $T_m^*$ grows linearly with the heat power and eventually
saturates to a maximum value ($T_m^*/T\simeq 2.8(1)$ in
Fig.~\ref{fig2}).  The linear behavior found in the limit $P\to 0$
(dashed line shown in Fig.~\ref{fig2}) can be expressed as
$P=\alpha(T_m^*/T-1)$ where $\alpha$ is a constant. We remark that this
linear dependence is unconventional. In Onsager theory, the heat flux or
current $j$ between two bodies at different temperatures is governed by
the Fourier law, $j=k\nabla T$ where $k$ is the thermal conductivity of
the system. The entropy production $P$, though, is proportional to
$(\nabla T)^2$. Accordingly, one might expect a quadratic (rather than
linear) dependence for the dissipated power $P$ with the difference
$(T_m^*-T)$, contrary to what is observed in Fig.~\ref{fig2}.  As the
dissipated power is always positive (as implied by the second law of
thermodynamics), the linear dependence suggests that the measured
temperature $T_m$ can never be lower than the bath temperature. This is
in agreement with our findings: we could not find values for the
parameters of the model such that $T_m$ was lower than $T$.  The
physical significance of the inequality, $T_m>T$, in the context of
nonequilibrium steady states, remains unclear to us~\footnote{In this
regard, it would be interesing to clarify whether this inequality is
somehow related to an inequality describing transitions between
nonequilibrium steady states put forward by Oono and
Paniconi$^{29}$ in the context of the steady-sate
thermodynamics. This inequality has been obtained as a particular case
of a more general nonequilibrium equality derived by Hatano and Sasa$^{30}$ and experimentally tested very
recently$^{31}$.}

\section{Discussion}

A few years ago, motivated by theoretical studies in spin glasses
and structural glass models,  Cugliandolo, Kurchan and Peliti$^8$ proposed to define nonequilibrium temperature
(NET) as the temperature that a cavity coupled to the system would
measure. The measured temperature could be determined by putting the
cavity in contact with its own thermal bath and modulating the
temperature of that bath until the net
heat power exchanged between the system and the cavity vanishes. In
the limit of large cavity frequencies, such a temperature was found to
coincide with the temperature obtained from the extended form of the
FDT. Yet, clear experimental evidence and simple experimental tests 
of this nonequilibrium temperature are yet to come.

In this paper, we have studied a simple nonequilibrium steady-state
system that can be described by a nonequilibrium temperature
(NET). It consists of a two-state system driven by an oscillating ac
field and coupled to a cavity. Under general conditions, we predict the
existence of a resonant peak for the NET at values of the driving
frequency $\omega$ that are equal to the resonant frequency of the
cavity $\omega_0$ (Figure~\ref{fig1}). For large values of $\omega_0$,
the measured temperature asymptotically converges to a well-defined
value that is a function of the dissipated power, allowing us to define a
constitutive equation for the steady-state system. The origin of this
resonance effect is mechanical; the frequency of the cavity $\omega_0$
resonates with the frequency of the ac field $\omega$. Although we did
not study it in detail, a similar resonant effect is predicted in the
limit of a purely overdamped cavity ($\omega_0\to 0$) when the
relaxational frequency of the cavity $\omega_{\rm dis}=1/\tau_{\rm
dis}=k/\gamma\sim \omega$. The origin of this resonance would then be
stochastic. This phenomenon would then fall in the category of
stochastic resonance phenomena,$^{32}$ an ubiquitous
behavior found in relaxational systems.

The results of Figures~\ref{fig1} and \ref{fig2} could be experimentally
tested in magnetic or dielectric systems. For instance, in the case of
dielectric systems, an oscillating voltage of intensity $V_0$ and
frequency $\omega$ could be applied between the metallic plates of a
capacitor containing a dielectric material with relaxational electric dipoles that
have a characteristic time scale $\tau_{\rm relax}$. The 
dipolar charge in the surface of the capacitor is
then coupled to an external electrical RLC circuit that acts as a cavity. The cavity has
characteristic resonant frequency $\omega_0=1/\sqrt{LC}$ and a damping
frequency $w_{\rm dis}=1/\tau_{\rm dis}=1/\sqrt{RC}$.  For a fixed
frequency of the ac electric voltage $\omega$ and the frequency of the
cavity $\omega_0$, one could then
determine the voltage noise density and the electric impedance
of the cavity from which one could obtain values for the measured $T_m$
by using the generalized Nyquist relation (eq \eq{eq5}).$^{10,11}$
This also would allow one to measure the dissipated power of the cavity
$P=V^2/|Z|$ ($Z$ being the electric impedance) as a function of $T_m$.
Reproducing the qualitative results of Figures~\ref{fig1} and \ref{fig2}
should be within reach with current experimental techniques.

The results of Figures~\ref{fig1} and \ref{fig2} could be also
experimentally tested in magnetic systems (e.g. nanoparticles or
paramagnetic systems) by coupling the ac magnetization signal to a resonant
electric cavity and measuring the voltage noise density at different
frequencies.  This technique has been used to measure
magnetostochastic resonance$^{32,33}$ in
ferrite-garnet films using the Faraday effect. Another tool could be
provided by transverse-field ac measurements$^{34}$.

Let us mention that results quite similar to those shown in
Figure~\ref{fig1} have been recently reported in the measurement of
effective temperatures in a colloidal suspension of aging Laponite by
using the nonequilibrium Einstein relation. The control parameter there
is the age of the system $t_w$ rather than the driving or probing
frequencies ($\omega$ and $\omega_0$ respectively).  In these
measurements the maximum of the temperature was interpreted as a
resonance between the probe frequency $\omega_0$ and the inverse of
the relaxation time scale of the suspension ($\tau_{\rm relax}$ in the
present case).

To conclude, given the elusive interpretation of the concept of
effective temperature and the formidable complexity of glassy systems,
it would be very instructive to carry out systematic research of
NETs in simple noninteracting systems driven to nonequilibrium steady
states.  A simple experiment such as the one that we have described should
not be difficult to carry out and would help us to better understand
what we can expect from noise measurements in more complex
nonequilibrium systems. In this respect, two-level systems are simple
model systems where theoretical predictions for the NETs could be
experimentally tested at both qualitative and quantitative levels.

{\bf Acknowledgments.} I am grateful to David Jou for a critical reading of
the manuscript. I also acknowledge support from the Spanish
Ministerio de Ciencia y Tecnolog\'{\i}a Grant BFM2001-3525, Generalitat
de Catalunya, the STIPCO network, and the SPHINX program.

\vspace{1cm}
{\bf\Large References}
\vspace{.5cm}

(1) Casas-Vazquez, J.; Jou, D. {\em Rep. Prog. Phys.} {\bf 2003}, {\em 66, 1937}

(2) De Groot, S. R.; Mazur, P. {\em Non Equilibrium Thermodynamics}; Dover Publications: New York, 1984 

(3) Vilar, J. M.; Rub\'{\i}, M. {\em Proc. Natl. Acad. Sci. USA} {\bf 201}, {\em 98, 11081}.

(4) Cugliandolo, L. F.; Kurchan, J. {\em Phys. Rev. Lett.} {\bf 2003}, {\em 71, 173}.

(5) For recent reviews see, L. F. Cugliandolo, Crisanti, A.; Ritort, F. {\em
J. Phys. A (Math. Gen.)} {\bf 2003}, {\em 36, R181}.

(6) Ritort, F In {\em Unifying Concepts in Granular Media
and Glasses}; Coniglio, A., et et al., Eds.; Elsevier: Amsterdam, 2004; p 129 (preprint {\bf cond-mat/0311370}).

(7) The robustness of the concept of
thermodynamic temperature has its roots in the statistical
interpretation of heat provided by statistical mechanics. According to
it, temperature is directly related to the gradient in the density of
energy levels $\rho(E)$, $1/T=\frac{\partial S(E)}{\partial E}$ where
$S(E)=k_B\log(\rho(E))$.  The statistical temperature is identical to
the thermodynamic temperature because two bodies at mutual equilibrium
reach a maximal entropy state compatible with the total energy
available.

(8) Cugliandolo, L. F.; Kurchan, J.; Peliti, L. {\em Phys. Rev. E} {\bf 1997}, {\em 55, 3898}

(9) Alba, M.; Hammann, J.; Ocio, M.; Refregier, P.; Bouchiat, H. {\em J. Appl. Phys.} {\bf 1987}, {\em 61, 3683}.

(10) Grigera, T. S.; Israeloff, N. E. {\em Phys. Rev. Lett.} {\bf 1999}, {\em 83, 5038}.

(11) Bellon, L.; Ciliberto, S.; Laroche, C. {\em 
Europhys. Lett.} {\bf 2001}, {\em 53, 511}; Buisson, L.; Bellon, L.; Ciliberto, D. {\em J. Phys.: Condens. Matter.)} {\bf 2003}, {\em 15, S1163};

(12) Herisson, D.; Ocio, M. {\em Phys. Rev. Lett.} {\bf 2002}, {\em 88, 257202}.

(13) D'Anna, G.; Mayor, P.; Barrat, A.; Loreto, V.; Nori, F.
 {\em Nature} {\bf 2002}, {\em 424, 1909}. 

(14)Ojha, R., P.; Lemieux, P. A.; Dixon, P. K. ; Liu, A.; Durian, D. J. {\em Nature} {\bf 2004}, {\em 427, 521}.

(15) Abou, B.; Gallet, F. {\it Probing a nonequilibrium
Einstein relation in an aging colloidal glass}, Preprint {\bf
cond-mat/0403561} to appear in Phys. Rev. Lett.

(16) Fielding, S. M.; Sollich, P. {\em Phys. Rev. Lett.} {\bf 2002}, {\em 88, 050603}.

(17) Ediger, M. D. {\em Annu. Rev. Phys. Chem.} {\bf 2000}, {\em 51, 99}.

(18) Sasa, S.; Tasaki, H. {\it Steady-state
thermodynamics for heat conduction}. Unpublished work (preprint {\bf cond-mat /0108365}).

(19) Hatano, T.; Jou, D. {\em Phys. Rev. E} {\bf 2003}, {\em 67, 026121}.

(20) Wernsdorfer, W. {\em Adv. Chem. Phys.} {\bf 2001}, {\em 118, 99}.

(21) Evans, E. {\em Annu. Rev. Biomol. Struct.} {\bf 2001}, {\em 30, 105}.

(22) Gonzalez-Miranda, J. M.; Tejada, J. {\em Phys. Rev. B} {\bf 1994}, {\em 49, 3867}.

(23) Wernsdorfer, W.; et al., {\em J. Magn. Magn. Mater.} {\bf 1995}, {\em 145, 33};  Wernsdorfer, W.; et al. {\em Phys. Rev. Lett.} {\bf 1997}, {\em 78, 1791}.

(24) Jamet, M. et al., {\em Phys. Rev. Lett.} {\bf 2001}, {\em 86, 4676}.

(25) Garriga, A.; Ritort, F. {\em Eur. Phys. J. B} {\bf 2001}, {\em 21, 115}.

(26) Ritort, F. {\em J. Stat. Mech: Theor. Exp.} {\bf 2004}, P10016 

(27) Hohenberg, P.; Shraiman, B. {\em Physica D} {\bf 1989}, {\em 37, 109}.

(28) Jou, D.; Casas-Vazquez, J.; Lebon, G. {\it
Extended Irreversible Thermodynamics}, 3rd ed.; Springer: Berlin, {\bf 2003}.

(29) Oono, Y.; Paniconi, M. {\em Prog. Theor. Phys. Suppl.} {\bf 1998}, {\em 130,29}.

(30) Hatano, T.; Sasa, S. {\em Phys. Rev. Lett.} {\bf 2001}, {\em 86,3463}.

(31) Trepagnier, E. H.; Jarzynski, C.; Ritort, F.; Crooks, G. E.; Bustamante, C. J.; Liphardt, J. {\em  Proc. Natl. Acad. Sci. USA} {\bf 2004}, {\em 101, 15038}.

(32) Gammaitoni, L.; et al., {\em Rev. Mod. Phys.} {\bf 1998}, {\em 70, 223}.

(33) Grigorenko, A. N.; et al, {\em J. Appl. Phys.} {\bf 1994}, {\em 76, 6335}.

(34) Spinu, L.; et al., {\em J. Appl. Phys.} {\bf 2000}, {\em 87, 5490}.

\end{document}